# Resistive Scaling in the Magnetic Helicity-driven Inverse Cascade

Jiyao Zhang[1] and Axel Brandenburg[2,3,4,5]
[1] Department of Mathematics, University of Pennsylvania, Pennsylvania, PA 19104, USA
[2] Nordita, KTH Royal Institute of Technology and Stockholm University, Hannes Alfvéns väg 12, SE-10691 Stockholm, Sweden
[3] The Oskar Klein Centre, Department of Astronomy, Stockholm University, AlbaNova, SE-10691 Stockholm, Sweden
[4] McWilliams Center for Cosmology & Department of Physics, Carnegie Mellon University, Pittsburgh, PA 15213, USA
[5] School of Natural Sciences and Medicine, Ilia State University, 3–5 Cholokashvili Avenue, 0194 Tbilisi, Georgia

## Abstract

The inverse cascade in MHD turbulence plays a crucial role in various astrophysical processes such as galaxy cluster formation, solar and stellar dynamo mechanisms, and the evolution of primordial magnetic fields in the early Universe. A standard numerical approach involves injecting magnetic helicity at intermediate length scales to generate a secondary, time-dependent spectral peak that gradually propagates toward larger scales. Previous simulations have already suggested a resistive dependence of inverse transfer rates and demonstrated the significant influence of magnetic helicity flux density $\epsilon_{\rm H}$ on this process. On dimensional grounds, we have for the spectral envelope $E_{\rm M}(k, t) \leqslant C_{\rm H}\epsilon_{\rm H}^{2/3}k^{-1}$, where $C_{\rm H}$ represents a potentially universal dimensionless coefficient analogous to the Kolmogorov constant. We present a summary of the 25 distinct simulations conducted with the PENCIL CODE, systematically varying the forcing wavenumber $k_f$, magnetic Prandtl number $\mathrm{Pr_M}$, grid resolution $N^3$, and Lundquist number Lu. We obtained $C_{\rm H}$ and corresponding error bars by calculating the compensated spectrum and investigated its dependence with Lu and $k_f$. For the $C_{\rm H}$–Lu relationship, we observe strong correlations with a power-law exponent around unity. In contrast, we find no significant correlation between $C_{\rm H}$ and $k_f$. We also present evidence for a nonresistive scaling of the form $E_{\rm M}(k, t) \leqslant C_{2/3}v_{\rm A}^2k^{-1}$, where $v_{\rm A}$ is the rms Alfvén speed and $C_{2/3} \approx 0.4$ for all of our runs.

## 1. Introduction

Turbulent flows involve a large range of length scales. Due to the presence of nonlinearities in the hydrodynamic equations, there can be energy transfer between different length scales. This energy transfer is typically local in wavenumber space, and therefore one tends to talk about an energy *cascade*. In ordinary three-dimensional hydrodynamic turbulence, energy flows from large to small scales, which is referred to as a direct or forward cascade. In the presence of magnetic fields, however, turbulence can behave very differently. In particular, there is the possibility of an inverse cascade if the magnetic field is helical. This was first explored by U. Frisch et al. (1975) and A. Pouquet et al. (1976), who associated the inverse cascade with the conservation of magnetic helicity.

The early work on inverse transfers in hydromagnetic turbulence is significant in the context of astrophysical magnetism. It was known that large-scale magnetic fields in stars and galaxies can be caused by cyclonic turbulence (E. N. Parker 1955, 1971). This means that the combination of radially inward directed gas density gradients and global rotation can cause negative kinetic helicity of the turbulence in the northern hemisphere and positive kinetic helicity in the southern hemisphere. Such flows render a nonmagnetic state unstable to small amplitude and large wavelength perturbations (H. K. Moffatt 1970, 1978). This was explained in terms of what is called the $\alpha$ effect (M. Steenbeck et al. 1966), where $\alpha$ is a pseudoscalar proportional to the negative kinetic helicity in the evolution equations for the mean magnetic field at sufficiently high conductivity.

The possibility of an inverse cascade of magnetic helicity toward larger scales was studied numerically by injecting magnetic helicity at intermediate length scales. This led to the emergence of a second, time-dependent peak in the magnetic energy spectrum that gradually propagated toward smaller wavenumbers, corresponding to progressively larger length scales (A. Pouquet et al. 1976). The first peak stays fixed and reflects the helicity injection wavenumber. A similar behavior can also be seen in simulations with finite kinetic helicity forcing (A. Brandenburg 2001) instead of the magnetic forcing.

While the simulations with kinetic forcing explained some important properties of astrophysical magnetism, they still have the problem of displaying a resistive decrease of the resulting mean magnetic field strengths with increasing magnetic Reynolds number (F. Del Sordo et al. 2013; F. Rincon 2021). Because of this, it still remains difficult to explain the large-scale magnetic field generation in astrophysically relevant systems at large magnetic Reynolds numbers. In that context, we mention the work of A. Brandenburg et al. (2002), who found evidence for a resistivity-dependent speed of the inverse transfer; see their Figure 11.

A possible solution to the problem of resistively limited large-scale magnetic field generation was thought to be the connected with magnetic helicity conservation within the domain. This was pointed out by A. V. Gruzinov & P. H. Diamond (1996), who argued that in the absence of magnetic helicity fluxes, as is the case in periodic domains, the magnetic helicity from the small-scale field leads to an adverse contribution to the $\alpha$ effect that is proportional to the current helicity density (A. Pouquet et al. 1976). These ideas emerged after the resistively slow saturation behavior of the magnetic field in the three-dimensional turbulence simulations of A. Brandenburg (2001) was understood to be a consequence of magnetic helicity conservation (E. G. Blackman & G. B. Field 2002); see A. Brandenburg et al. (2002).

Significant effort has gone into the study of magnetic helicity fluxes (E. T. Vishniac & J. Cho 2001; K. Subramanian & A. Brandenburg 2004, 2006; A. Hubbard & A. Brandenburg 2011, 2012; F. Del Sordo et al. 2013; F. Rincon 2021). However, not only the saturation magnetic field strength but also the magnetic helicity fluxes themselves have continued to depend on the magnetic Reynolds number until the present day. In the recent work of A. Brandenburg & E. T. Vishniac (2025), it was shown that the spatial magnetic helicity fluxes between regions of different magnetic helicity density can be equal to the spectral ones from small to large length scales. This motivates a fresh look at the dependence of the speed of magnetic helicity fluxes on the magnetic Reynolds number.

A resistive dependence of the speed of inverse transfer in the inertial range of magnetically forced turbulence has previously been seen in the simulations of A. Brandenburg et al. (2002). This was surprising, because in turbulence, the microphysical viscosity and resistivity were thought to not play a role and should not affect the turbulence as a whole. To reexamine this possibility, it is useful to adopt a more idealized setting where magnetic helicity is injected directly at intermediate length scales, just as was done in the original work of A. Pouquet et al. (1976). Similar models have also been considered on other occasions (S. K. Malapaka & W.-C. Müller 2013).

One of the key points of the present investigation is the analysis of the dimensionally motivated law for the spectral magnetic energy evolution. One may argue that the main physical process governing the system is the magnetic helicity flux density $\epsilon_{\rm H}$, which has units of magnetic helicity density per unit time. Assuming that the magnetic field is characterized by the Alfvén velocity $v_{\rm A} = B_{\rm rms}/\sqrt{\mu_0\rho_0}$, where $B_{\rm rms}$ is the rms magnetic field, $\mu_0$ is the vacuum permeability, and $\rho_0$ is the background density, the magnetic helicity density, divided by $\mu_0\rho_0$, has units of $v_{\rm A}^2\xi_{\rm M}$, where $\xi_{\rm M}$ is a characteristic magnetic length scale, so the units are $\rm cm^3\,s^{-2}$. The units of the magnetic helicity flux are therefore $\rm cm^3\,s^{-3}$. We employ the magnetic energy spectrum defined such that $\int E_{\rm M}(k,t)\,dk = v_{\rm A}^2/2$, where $k$ is the wavenumber. Since $k$ has units of $\rm cm^{-1}$, the units of $E_{\rm M}(k,t)$ are $\rm cm^3\,s^{-2}$. Expressing $E_{\rm M}(k,t)$ as powers $a$ and $b$ of $\epsilon_{\rm H}$ and $k$, respectively, we have $E_{\rm M}(k,t) \propto \epsilon_{\rm H}^a k^b$. On dimensional grounds, we have $a = 2/3$ and $b = -1$, i.e.,

$$E_{\rm M}(k,t) = C_{\rm H}\epsilon_{\rm H}^{2/3}k^{-1}, \tag{1}$$

where $C_{\rm H}$ is a nondimensional coefficient. Assuming that this is indeed the relevant phenomenology, it is in principle possible that $C_{\rm H}$ is a universal constant, just like the Kolmogorov constant, which is the nondimensional constant in the kinetic energy spectrum in terms of a 2/3 power of the kinetic energy flux and a −5/3 power of the wavenumber. Alternatively, it is possible that $C_{\rm H}$ is different from case to case. This will be the possibility favored by the present simulations.

It should be pointed out that there is another possible phenomenology for a $k^{-1}$ spectrum, which assumes the presence of a large-scale magnetic field with Alfvén speed $v_{\rm A}$, so $E_{\rm M}(k) \propto v_{\rm A}^2 k^{-1}$ (A. A. Ruzmaikin & A. M. Shukurov 1982; N. Kleeorin & I. Rogachevskii 1994). This alternative is independent of the presence of magnetic helicity and may therefore not be relevant to us, because there would be no inverse cascade without net helicity. Also, in our case, the $k^{-1}$ power law describes the envelope of the inversely cascading peak of the spectrum rather than a continuous $k^{-1}$ spectrum over an extended range. The latter is expected when there is instead an already existing large-scale magnetic field characterized by $v_{\rm A}$; see Equation (31) of N. Kleeorin & I. Rogachevskii (1994). This is why those authors quoted the simulations of the preprint of A. Brandenburg et al. (1996); see their Figures 17 and 18.

The present paper is organized as follows. In Section 2, we begin by presenting the model, characteristic indicators, and initial conditions for direct numerical simulations of the forced helical MHD equations. In Section 3, we present the results derived from our numerical simulations, offering insights into the dependency of $C_{\rm H}$ with respect to Lundquist number Lu and forcing wavenumber $k_f$. Finally, we conclude our findings and discuss extending investigations in Section 4.

## 2. Numerical Simulations

### 2.1. Governing Equations in Helically Forced MHD

In this section, we consider MHD equations with an isothermal equation of state in a periodic domain with helical magnetic forcing. An isothermal equation of state is characterized by gas pressure $p$ proportional to the gas density $\rho$ with $p = \rho c_s^2$, where $c_s$ is a constant isothermal sound speed. To guarantee solenoidality, the magnetic field could be expressed in magnetic vector potential $\boldsymbol{A}$, i.e., $\boldsymbol{B} = \boldsymbol{\nabla}\times\boldsymbol{A}$. We solve the governing equations with the evolution equations for $\boldsymbol{A}$ and the velocity field $\boldsymbol{u}$ as follows:

$$\frac{\partial \boldsymbol{A}}{\partial t} = \boldsymbol{u}\times\boldsymbol{B} - \eta\mu_0\boldsymbol{J} + \boldsymbol{\mathcal{E}}_{\rm ext}, \tag{2}$$

$$\frac{{\rm D}\boldsymbol{u}}{{\rm D}t} = -c_{\rm s}^2\boldsymbol{\nabla}\ln\rho + \frac{1}{\rho}[\boldsymbol{J}\times\boldsymbol{B} + \boldsymbol{\nabla}\cdot(2\nu\rho\boldsymbol{\mathsf{S}})], \tag{3}$$

$$\frac{{\rm D}\ln\rho}{{\rm D}t} = -\boldsymbol{\nabla}\cdot\boldsymbol{u}, \tag{4}$$

where ${\rm D}/{\rm D}t = \partial/\partial t + \boldsymbol{u}\cdot\boldsymbol{\nabla}$ is the advective derivative, $\boldsymbol{\mathcal{E}}_{\rm ext}$ is the external forcing function, $\boldsymbol{J} = \boldsymbol{\nabla}\times\boldsymbol{B}/\mu_0$ is the current density, $\eta$ is the magnetic diffusivity, $\nu$ is the kinematic viscosity, and $\boldsymbol{\mathsf{S}}$ is the traceless rate-of-strain tensor with the following components:

$$\mathsf{S}_{ij} = \frac{1}{2}(\partial_i u_j + \partial_j u_i) - \frac{1}{3}\delta_{ij}\boldsymbol{\nabla}\cdot\boldsymbol{u}. \tag{5}$$

Owing to the absence of boundaries, and using volume averages indicated by angle brackets, the magnetic helicity equation is then given by

$$\frac{d}{dt}\langle\boldsymbol{A}\cdot\boldsymbol{B}\rangle = -2\eta\mu_0\langle\boldsymbol{J}\cdot\boldsymbol{B}\rangle + 2\langle\boldsymbol{\mathcal{E}}_{\rm ext}\cdot\boldsymbol{B}\rangle, \tag{6}$$

where the first term on the right-hand side quantifies the resistive losses and the second term the magnetic helicity injection through the forcing function. If there were a statistically steady state, the two terms on the right-hand side of Equation (6) should be equal. In addition, owing to the stochastic nature of the forcing, the determination of $2\langle \boldsymbol{\mathcal{E}}_{\rm ext} \cdot \boldsymbol{B}\rangle$ is less accurate. Therefore, in our numerical analysis, we estimate the magnetic helicity flux through the dissipative term, i.e., $\epsilon_{\rm H} = 2\eta\mu_0\langle \boldsymbol{J} \cdot \boldsymbol{B}\rangle$.

### 2.2. The Model

We solve Equations (2)–(4) with periodic boundary conditions using the PENCIL CODE, which employs sixth-order finite differences and a third-order accurate time-stepping scheme. We compare runs with different resolutions using up to $N^3 = 1024^3$ mesh points. We use a fifth-order upwind derivative operator for the advection term (W. Dobler et al. 2006) to damp spatial oscillations at the Nyquist wavenumber. Each simulation is further characterized by the Lundquist number Lu, which is defined as

$$\mathrm{Lu} = \frac{B_{\rm rms}}{\eta k_{\rm p}} = \frac{v_{\rm A}\xi_{\rm M}}{\eta}, \tag{7}$$

where $k_{\rm p}$ is the peak forcing wavenumber of the spectrum, $v_{\rm A} = B_{\rm rms}/\sqrt{\mu_0/\rho_0}$ is the Alfvén speed based on the rms magnetic field, and $\xi_{\rm M} = 1/k_{\rm p}$ is a magnetic correlation length, which is also characterized by

$$\xi_{\rm M}(t) = \frac{\int k^{-1}E_{\rm M}(k)\,dk}{\int E_{\rm M}(k)\,dk}. \tag{8}$$

The simulations are further characterized by the fluid and magnetic Reynolds numbers,

$$\mathrm{Re} = \frac{u_{\rm rms}\xi_{\rm M}}{\nu} = \frac{u_{\rm rms}}{\nu k_{\rm p}}, \quad \mathrm{Re}_{\rm M} = \frac{u_{\rm rms}\xi_{\rm M}}{\eta} = \frac{u_{\rm rms}}{\eta k_{\rm p}}, \tag{9}$$

where $u_{\rm rms}$ is the rms value of the resulting velocity field and the magnetic Prandtl number is given by $\mathrm{Pr}_{\rm M} = \nu/\eta = \mathrm{Re}_{\rm M}/\mathrm{Re}$. The forcing function $\boldsymbol{\mathcal{E}}_{\rm ext}$ in Equation (2) is randomly chosen and $\delta$-correlated in time, defined as

$$\boldsymbol{f}(x, t) = \mathrm{Re}\{f_0 c_s(|\boldsymbol{k}|c_s/\delta t)^{1/2}\boldsymbol{f}_{\boldsymbol{k}(t)}\, e^{i[\boldsymbol{k}(t)\cdot\boldsymbol{x}+\phi(t)]}\}, \tag{10}$$

where $f_0$ is a nondimensional forcing amplitude, $\delta t$ is the length of the time step, $-\pi < \phi(t) < \pi$ is a random phase, and $\boldsymbol{k}(t)$ is randomly chosen from a pregenerated set of wavevectors in a narrow band of width $\delta k$ around a given forcing wavenumber with an average value $k_f$, i.e.,

$$k_{\rm f} - \delta k/2 \leqslant |\boldsymbol{k}(t)| < k_{\rm f} + \delta k/2. \tag{11}$$

In all cases, the amplitude of the forcing function is $f_0 = 0.01$, which results in a Mach number $u_{\rm rms}/c_s$ of around 0.05. Transverse helical waves are produced via (A. Brandenburg & K. Subramanian 2005)

$$\boldsymbol{f}_{\boldsymbol{k}} = \mathsf{R} \cdot \boldsymbol{f}_{\boldsymbol{k}}^{\rm nohel}, \quad \mathsf{R}_{ij} = \frac{\delta_{ij} - i\sigma\epsilon_{ijk}\hat{k}_k}{\sqrt{1+\sigma^2}}, \tag{12}$$

where $\sigma$ is a measure of the helicity of the forcing. In our case, we keep $\sigma = 1$ for positive maximum helicity of the forcing function, and

$$\boldsymbol{f}_{\boldsymbol{k}}^{\rm nohel} = (\boldsymbol{k} \times \hat{\boldsymbol{e}})/\sqrt{k^2 - (\boldsymbol{k}\cdot\hat{\boldsymbol{e}})^2} \tag{13}$$

is a nonhelical forcing function, where $\hat{\boldsymbol{e}}$ is an arbitrary unit vector that is not aligned with $\boldsymbol{k}$.

Our initial conditions are $\boldsymbol{A} = \boldsymbol{u} = \ln(\rho/\rho_0) = 0$, where $\rho_0$ is the mean density, which is a constant owing to mass conservation and the use of periodic boundary conditions. Starting with the first time step, $\boldsymbol{A}(\boldsymbol{x}, t)$ begins to evolve away from zero. The resulting Lorentz force $\boldsymbol{J} \times \boldsymbol{B}$ then drives $\boldsymbol{u}$ away from zero, and finally, finite compressions with $\boldsymbol{\nabla} \cdot \boldsymbol{u} \Leftrightarrow 0$ drive $\ln(\rho/\rho_0)$ away from zero.

## 3. Results

In Table 1, we present a summary of the runs discussed in this paper. Figure 1 illustrates the inverse cascade process using simulation A1, showing energy spectra at different time points with red dots marking the calculated spectral peaks, which clearly reveal the characteristic envelope of the inverse cascade evolution. Note, the spectrum peak on the right side is caused by the injection of helical forcing, which occurs at $k_f$.

We separated our simulations into subsets so that we could examine the dependence of $C_{\rm H}$ with respect to the various variables (e.g., $k_f$, $\mathrm{Pr}_{\rm M}$, $N^3$, and Lu). Our primary focus is to cover a range of values of Lu from 5 to 54, but we also altered Re and $\mathrm{Re}_{\rm M}$ so as to obtain a range of $\mathrm{Pr}_{\rm M}$ from 2/5 to 2 to examine whether it plays a role in the inverse cascade process of evolving helical MHD. We also did simulations for multiple resolutions to measure the uncertainty caused by mesh resolution.

First of all, we examine the helicity dissipation decay of each run. In Figure 2, we plot $\epsilon_{\rm H}$ versus time for simulation subset A as an illustration. We see that $\epsilon_{\rm H}$ levels off at late times, but we find the asymptotic values tend to decrease with decreasing values of $\eta$. We can make the curves approximately overlap by scaling them with $(\eta/\eta_{-4})^{0.6}$ (Figure 2(b)). Here, we have chosen to normalize by $\eta_{-4} = 10^{-4}$, the value of one for A1. It is clear that the magnetic energy spectrum does not follow a universal decay law, and that the magnetic helicity dissipation is mostly controlled by $\eta\langle \boldsymbol{J} \cdot \boldsymbol{B}\rangle$ and the system is within the same range of physical control parameters.

In all cases, the nondimensional coefficient $C_{\rm H}$ in Equation (1) can be estimated by fitting a power law to the spectrum peak at selected iterative time steps during inverse cascade (U. Frisch et al. 1975). The position of the spectral peak is calculated using $k_{\rm p} = 1/\xi_{\rm M}$. Note that here and in Equation (8), we have chosen to define $\xi_M$ without a $2\pi$ factor.

During the initial phase of each simulation, nonlinear interactions remain underdeveloped, and the energy spectrum is predominantly influenced by initial conditions or external forcing. To ensure that energy transfer to larger scales operates efficiently, we exclude the early times from our analysis. Similarly, at later stages when energy accumulates at the largest available scales, there are constraints imposed by finite domain size, potentially leading to artificial damping of large-scale modes through numerical viscosity or boundary effects. Consequently, we also need to exclude time steps occurring after the cessation of efficient energy transfer. Manual selection of the intermediate stage where energy cascade dominates introduces potential complications and subjective bias. In practice, we implement a systematic logarithmic sampling strategy, retaining snapshots at times corresponding to powers of 2 in addition to the initial time step. We systematically vary the starting time step from $t = 0$ and fit the corresponding spectral peaks using Equation (1). The configuration yielding the highest coefficient of determination ($R^2$) is selected to determine the final value of $C_\mathrm{H}$. $R^2$ is defined as

$$R^2 = 1 - \frac{\sum_i (E_\mathrm{M}(k) - E_\mathrm{M}^\mathrm{fit}(k))^2}{\sum_i (E_\mathrm{M}(k) - \langle E_\mathrm{M}(k) \rangle)^2}, \quad (14)$$

where $E_\mathrm{M}(k_i)$ are the numerical data points retrieved from each simulation, $E_\mathrm{M}^\mathrm{fit}(k_i)$ the corresponding fit values, and $\langle E_\mathrm{M}(k_i) \rangle$ their mean. $R^2$ measures the fraction of the variance in the data explained by the fit, with $R^2 = 1$ corresponding to a perfect fit.

**Table 1**
Overview of Simulation Runs in This Work

| Run | $\eta k_1/c_s$ | Re | $\mathrm{Re_M}$ | $\mathrm{Pr_M}$ | $k_f$ | $\epsilon_\mathrm{H}$ | Lu | $B_\mathrm{rms}/c_s$ | $C_\mathrm{H}$ | $C_{2/3}$ | $R^2$ | $N^3$ |
|---|---|---|---|---|---|---|---|---|---|---|---|---|
| A1 | $1.0 \times 10^{-4}$ | 27 | 27 | 1 | 80 | $8.48 \times 10^{-4}$ | 54 | 0.429 | 9.01 | 0.47 | 0.99 | $1024^3$ |
| A2 | $2.0 \times 10^{-4}$ | 12 | 12 | 1 | 80 | $1.36 \times 10^{-3}$ | 24 | 0.382 | 4.65 | 0.43 | 0.99 | $1024^3$ |
| A3 | $5.0 \times 10^{-4}$ | 4 | 4 | 1 | 80 | $1.89 \times 10^{-3}$ | 9 | 0.367 | 2.03 | 0.36 | 0.96 | $1024^3$ |
| B1 | $1.0 \times 10^{-4}$ | 12 | 25 | 2 | 80 | $8.39 \times 10^{-4}$ | 53 | 0.423 | 8.87 | 0.47 | 0.99 | $1024^3$ |
| B2 | $1.0 \times 10^{-4}$ | 11 | 23 | 2 | 100 | $1.27 \times 10^{-3}$ | 50 | 0.495 | 9.32 | 0.47 | 0.99 | $1024^3$ |
| B3 | $1.0 \times 10^{-4}$ | 10 | 21 | 2 | 120 | $4.19 \times 10^{-2}$ | 46 | 0.553 | 8.24 | 0.47 | 0.99 | $1024^3$ |
| B4 | $1.0 \times 10^{-4}$ | 25 | 25 | 1 | 200 | $3.44 \times 10^{-2}$ | 38 | 0.758 | 17.41 | 0.57 | 0.96 | $1024^3$ |
| C1 | $2.0 \times 10^{-4}$ | 12 | 12 | 1 | 80 | $1.42 \times 10^{-3}$ | 23 | 0.371 | 4.38 | 0.44 | 0.99 | $1024^3$ |
| C2 | $2.0 \times 10^{-4}$ | 11 | 11 | 1 | 100 | $2.19 \times 10^{-3}$ | 22 | 0.438 | 4.64 | 0.44 | 0.99 | $1024^3$ |
| C3 | $2.0 \times 10^{-4}$ | 10 | 10 | 1 | 120 | $3.06 \times 10^{-3}$ | 21 | 0.499 | 4.59 | 0.46 | 0.98 | $1024^3$ |
| C4 | $2.0 \times 10^{-4}$ | 8 | 8 | 1 | 200 | $7.64 \times 10^{-3}$ | 17 | 0.683 | 6.06 | 0.48 | 0.97 | $1024^3$ |
| D1 | $5.0 \times 10^{-4}$ | 3 | 3 | 1 | 80 | $2.28 \times 10^{-3}$ | 7 | 0.275 | 1.57 | 0.32 | 0.95 | $1024^3$ |
| D2 | $5.0 \times 10^{-4}$ | 3 | 3 | 1 | 100 | $2.93 \times 10^{-3}$ | 6 | 0.313 | 1.73 | 0.34 | 0.98 | $1024^3$ |
| D3 | $5.0 \times 10^{-4}$ | 2 | 2 | 1 | 120 | $6.29 \times 10^{-3}$ | 5 | 0.293 | 1.85 | 0.40 | 0.92 | $1024^3$ |
| D4 | $5.0 \times 10^{-4}$ | 2 | 2 | 1 | 200 | $1.27 \times 10^{-2}$ | 6 | 0.572 | 1.99 | 0.44 | 0.93 | $1024^3$ |
| E1 | $2.0 \times 10^{-4}$ | 11 | 11 | 1 | 80 | $8.48 \times 10^{-4}$ | 26 | 0.416 | 5.43 | 0.42 | 0.98 | $512^3$ |
| E2 | $3.0 \times 10^{-4}$ | 11 | 7 | 2/3 | 80 | $1.36 \times 10^{-3}$ | 15 | 0.353 | 3.12 | 0.40 | 0.98 | $512^3$ |
| E3 | $4.0 \times 10^{-4}$ | 10 | 5 | 1/2 | 80 | $1.89 \times 10^{-3}$ | 10 | 0.322 | 2.40 | 0.39 | 0.99 | $512^3$ |
| E4 | $5.0 \times 10^{-4}$ | 10 | 4 | 2/5 | 80 | $8.48 \times 10^{-4}$ | 7 | 0.291 | 1.76 | 0.38 | 0.98 | $512^3$ |
| E5 | $2.0 \times 10^{-4}$ | 10 | 10 | 1 | 100 | $1.88 \times 10^{-3}$ | 24 | 0.477 | 5.45 | 0.48 | 0.98 | $512^3$ |
| E6 | $2.0 \times 10^{-4}$ | 9 | 9 | 1 | 120 | $2.66 \times 10^{-3}$ | 22 | 0.524 | 5.24 | 0.50 | 0.98 | $512^3$ |
| F1 | $2.0 \times 10^{-4}$ | 12 | 12 | 1 | 80 | $8.37 \times 10^{-4}$ | 29 | 0.470 | 7.03 | 0.46 | 0.97 | $256^3$ |
| F2 | $3.0 \times 10^{-4}$ | 11 | 7 | 2/3 | 80 | $1.14 \times 10^{-3}$ | 18 | 0.430 | 4.79 | 0.39 | 0.98 | $256^3$ |
| F3 | $4.0 \times 10^{-4}$ | 10 | 5 | 1/2 | 80 | $1.39 \times 10^{-3}$ | 12 | 0.385 | 3.91 | 0.35 | 0.98 | $256^3$ |
| F4 | $5.0 \times 10^{-4}$ | 11 | 4 | 2/5 | 80 | $1.64 \times 10^{-3}$ | 9 | 0.369 | 2.05 | 0.37 | 0.98 | $256^3$ |

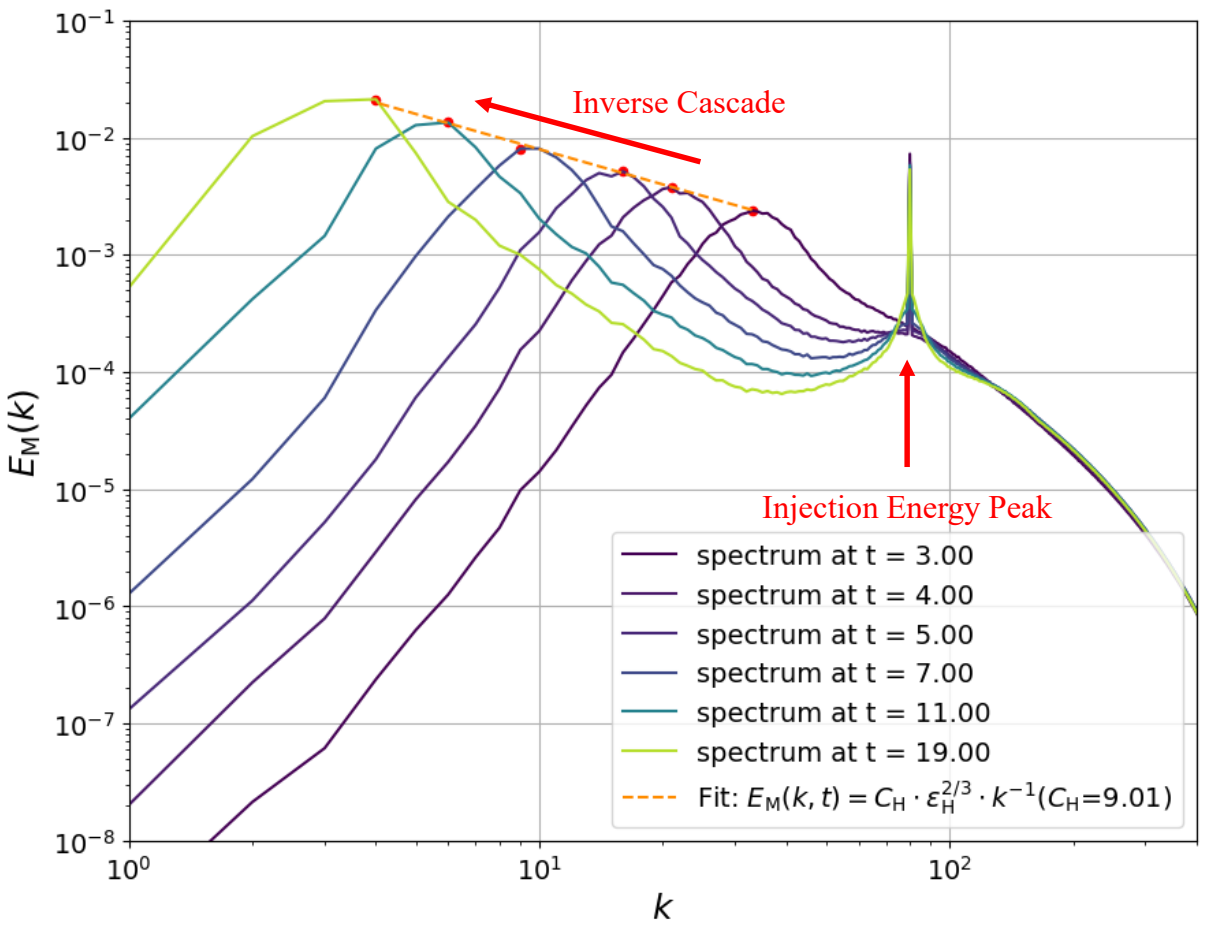

**Figure 1.** An illustration of estimating $C_\mathrm{H}$ using simulation A1. The red solid dots refer to the energy spectrum peak at each time step. The orange dashed line refers to the fitted curve from Equation (1) with $R^2 = 0.99$.

We further illustrate the validation of $C_\mathrm{H}$ by plotting the compensated magnetic energy spectra

$$E_\mathrm{M}^\mathrm{comp}(k) = k \epsilon_\mathrm{H}^{-2/3} E_\mathrm{M}(k). \quad (15)$$

This is shown in Figure 3 for simulation subset A. We see that the second peak evolves underneath an approximately flat envelope, whose value allows us to read off directly the value of $C_\mathrm{H}$ in each spectrum. We further obtained an error bar for each simulation run by setting the lowest and highest compensated spectrum peaks as the upper and lower bounds, respectively.

In summary, we find a clear correspondence between simulation precision and grid resolution. Simulations performed on the $256^3$ grid exhibit the largest uncertainty, with an average lower and upper error of 1.25 and 2.95 in the value of the exponent. Increasing the resolution to $512^3$ reduces the uncertainties to average lower and upper errors of 0.87 and 1.52, respectively, while the highest resolution, $1024^3$, yields the smallest uncertainties, with average lower and upper errors of 0.64 and 1.33. For the fitted $C_\mathrm{H}$ estimates, $R^2$ is consistent across grid resolutions. Specifically, the mean fitted $R^2$ values are 0.97, 0.98, and 0.98 for $1024^3$, $512^3$, and $256^3$ grids, respectively, with corresponding standard deviations of 0.018, 0.005, and 0.005. This indicates that, despite resolution-dependent differences in uncertainty, the quality of the fits remains comparable across all grids. Nevertheless, finer grids do provide more precise estimates of the fitted $C_\mathrm{H}$ and are therefore more reliable for the subsequent analysis. For each distinct fitted $C_\mathrm{H}$ value and corresponding $R^2$, we refer to Table 1.

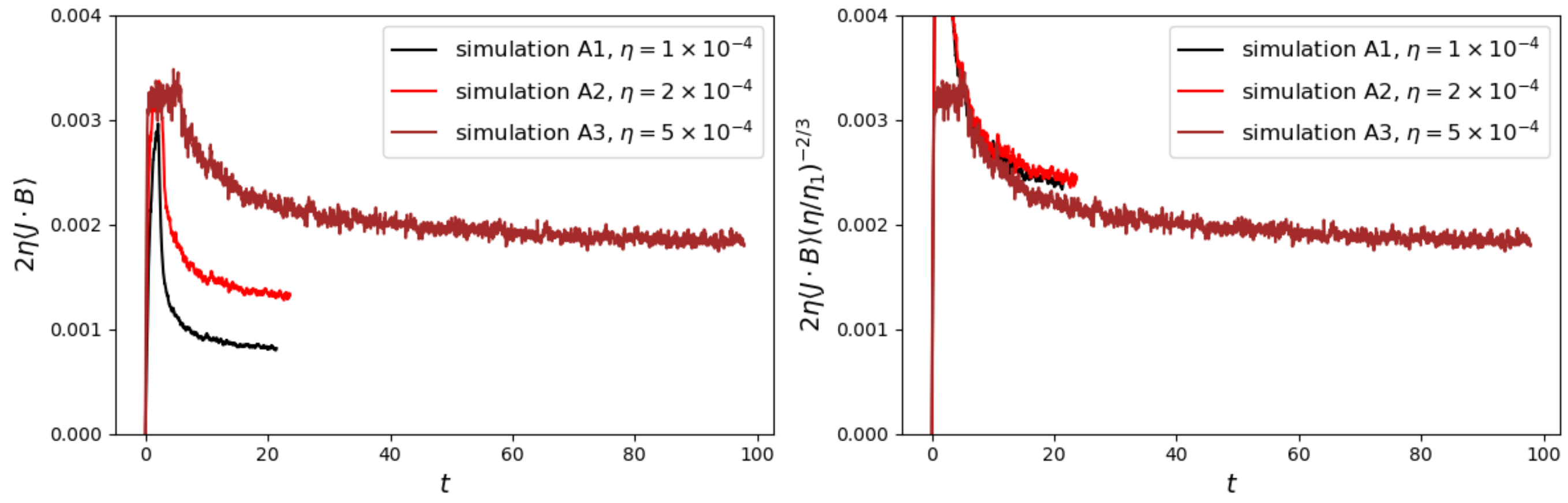

**Figure 2.** Helicity dissipation decay with respect to time for simulation runs A1 (black), A2 (red), and A3 (brown). Left panel: estimation of $\epsilon_{\rm H}$ by $2\eta\mu_0\langle \boldsymbol{J} \cdot \boldsymbol{B}\rangle$. Right panel: scaled $\epsilon_{\rm H}$ with $(\eta/\eta_1)^{-2/3}$ to make decay overlap with each run.

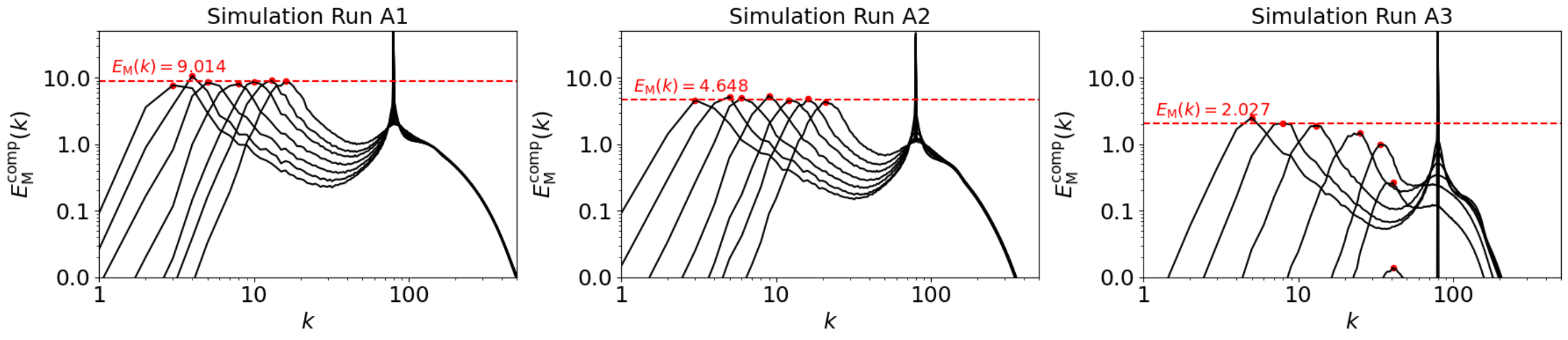

**Figure 3.** Compensated magnetic energy spectra for simulation runs A1 (left), A2 (middle), and A3 (right) at $t = 5, 6, 7, 9, 12, 16, 21$ (from right to left). The red dots illustrate the spectral peak at each time step. The red dashed horizontal line refers to approximated $C_{\rm H}$ ($R^2 =$ 0.99, 0.99, and 0.96, respectively).

Next, with the fitted $C_{\rm H}$ in each run, we examine the dependence of $C_{\rm H}$ with respect to Lu. We enabled larger forcing numbers to generate simulations with larger Lundquist number Lu. In Figure 4, we show the $C_{\rm H}$ dependence of Lu for the mesh points $N^3 = 256^3$, $512^3$, and $1024^3$. The simulations show that the ratio $C_{\rm H}$ scales with Lu,

$$C_{\rm H} \propto p\ {\rm Lu}^{q}, \tag{16}$$

but the exponent is not always the same. For $N^3 = 256$, we find $q \approx 1.11$ for both small and large values of Lu, while for $N^3 = 512$ and $N^3 = 1024$, we find $q \approx 1.18$ and $q \approx 0.69$.

Substituting Lu into the energy spectrum yields

$$E_{\rm M}(k,\, t) = C_q v_{\rm A}^{q+4/3}\eta^{2/3-q}k_f^{2/3-q}k^{-1}, \tag{17}$$

where $C_q$ is a nondimensional coefficient. Given that the exponential factor consistently approaches unity, we impose the constraint $q = 1$ and perform a single-parameter fit for the coefficient $p$. With this constraint applied to the $N = 1024$ simulation data, we obtain $p \approx 0.24$ with a coefficient of determination $R^2 = 0.82$ (Figure 4). The lower-resolution cases demonstrate improved agreement with this scaling law, yielding $p \approx 0.23$ for both the $N = 256$ and $N = 512$ configurations, with corresponding $R^2$ values of 0.90 and 0.98, respectively. This enhanced correlation at lower resolutions suggests that finite-size effects may influence the scaling behavior at higher grid densities.

Although the choice of $q$ does not affect the dimensional ground, the special choice of $q = 2/3$ would also yield a meaningful result, as discussed earlier in the introduction. For each run, the values of $C_{2/3}$ are given in Table 1. They are found to be around 0.4. Note that $C_{2/3}$ is related to $C_{\rm H}$. Specifically, we have, on average, $v_{\rm A}^2 C_{2/3} = \epsilon_{\rm H}^{2/3} C_{\rm H}$. Toward the end of the run, however, this relation would give too small values for $C_{2/3}$, because $v_{\rm A}$ is slowly increasing, while $\epsilon_{\rm H}$ is approximately constant.

We also perform a single-parameter fit for the coefficient $p$. We apply a similar single-parameter fitting procedure to determine the coefficient $p$. For $N = 1024$ simulation data, we obtain $p \approx 0.65$ with a coefficient of determination $R^2 = 0.84$ (Figure 4). The lower-resolution cases yield a coefficient of $p \approx 0.66$ and $p \approx 0.58$ for the $N = 256$ and $N = 512$ configurations, with corresponding $R^2$ values of 0.81 and 0.90, respectively.

We should emphasize that there is no strong physical reason to favor a particular value of $q$. As explained above, the case $q = 2/3$ can be motivated by arguing that a magnetic field of successively larger scale is actually present, as evidenced by the pronounced peak of the spectrum traveling to smaller $k$. For the case $q = 1$, on the other hand, which corresponds to a linear dependence of $C_{\rm H}$ on Lu, we are not aware of any physical argument. In particular, if $q = 1$ were to be inserted into Equation (17), the resulting expression would just look more complicated.

One might be worried that these results are artifacts of the Lu still being too small and not yet in the asymptotic regime in which a true Lu independence might be expected. However, comparing the energy spectra in at least some of the cases with larger forcing wavenumbers indicates that there is indeed a range of Lu = 4 to Lu = 50 in which there is an approximate $p$ scaling. On the other hand, however, we notice that with a larger forcing wavenumber, the estimated $C_H$ tends to be larger with the same Lu condition, i.e., they tend to produce simulation points at the upper left in Figure 4. This may also be regarded as evidence that none of the present simulations are yet in the truly asymptotic regime. Therefore, even higher-resolution simulations at larger Lundquist numbers remain essential.

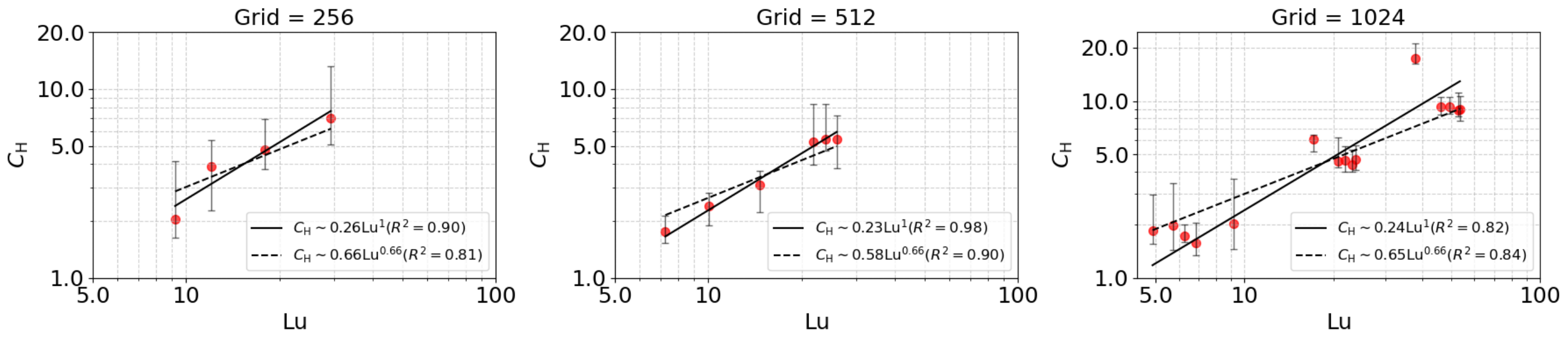

**Figure 4.** Dependence of the $C_H$ on Lundquist number Lu for $N = 256$ (left), 512 (middle), and 1024 (right) simulations of power 1 (solid black line) and power 2/3 (dashed black line). Each point refers to a distinct simulation run. The solid vertical lines refer to error bar estimates.

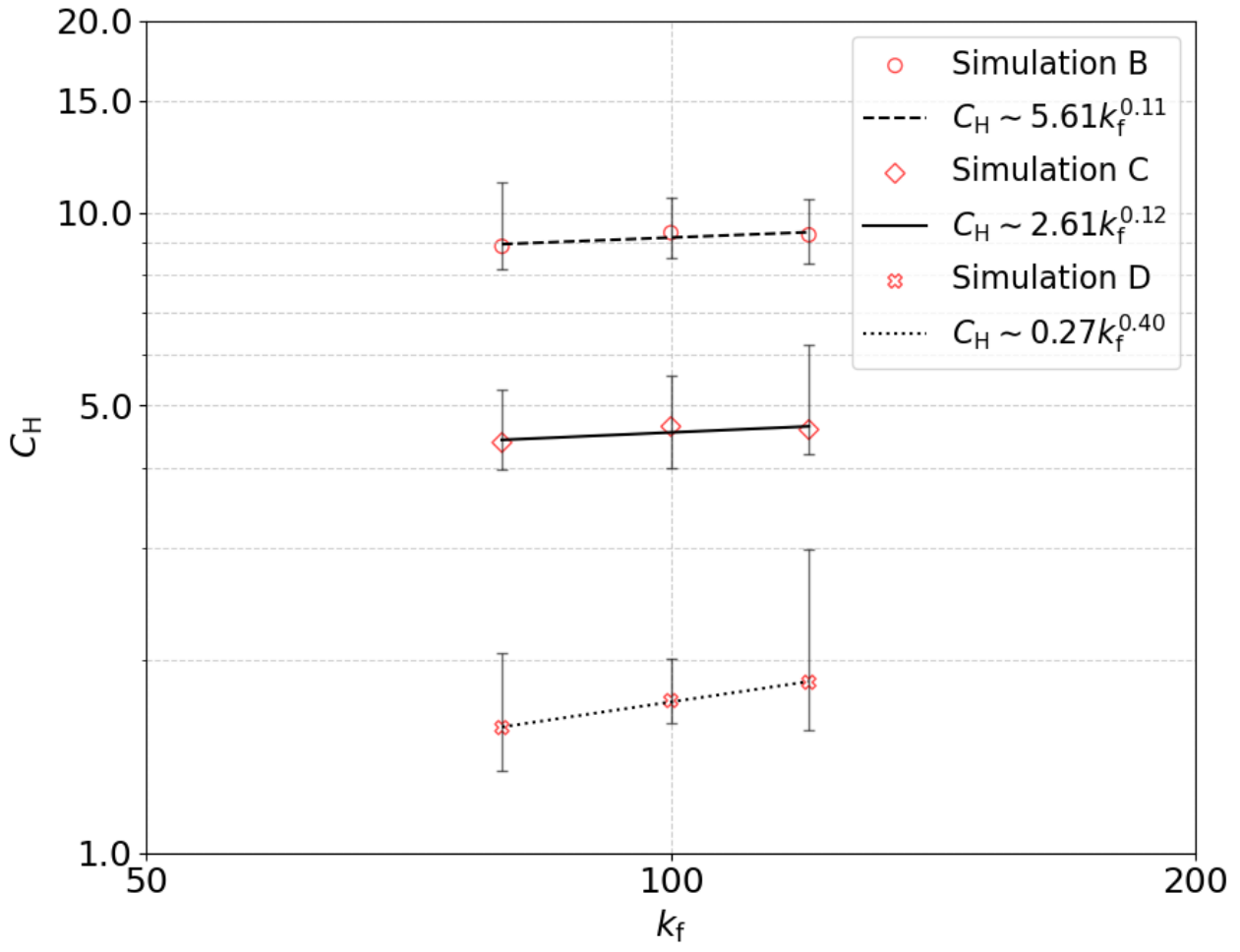

**Figure 5.** Dependence of the $C_H$ on forcing wavenumber $k_f$ for simulation subset B (red open circles and dashed black line), C (red open diamonds and solid black line), and D (red open crosses and dotted black line). The solid vertical lines refer to error bar estimates.

Next, we examine the dependence of $C_H$ with respect to the forcing wavenumber $k_f$. Subsets B, C, and D exhibit consistency with other characteristic indicators such as Lu and $\text{Pr}_M$, making them suitable for investigating the effects of varying $k_f$. We exclude runs in the subset with $k_f/k_1 = 200$ since they tend to be less accurate, i.e., they yield a relatively low $R^2$ and wider error bars. Similarly, to determine the functional dependence, we fit the relationship $C_H \propto p\, k_f^q$. Figure 5 shows the fitted curves for all three subsets. The exponent $q$ remains small across all cases, with proportionality coefficients of 5.61, 2.61, and 0.27 for subsets B, C, and D, respectively. While the fitted curves achieve relatively high $R^2$ scores, the corresponding $p$-values, i.e., the probability of observing such data under the null hypothesis of no relationship, are elevated to 0.41, 0.39, and 0.03. Since these values exceed the two-sided significance threshold of 0.025 (2.5%), the relationships are statistically insignificant, although the case with $p = 0.03$ (subset D) is only marginally above this threshold. Therefore, we find no robust evidence for a clear dependence between $C_H$ and $k_f$.

This absence of a dependence between $C_H$ and $k_f$ is expected, as $C_H$ characterizes the efficiency of helicity energy transfer among scales and is therefore expected to be primarily controlled by global parameters such as Lu and $\text{Pr}_M$. Once such global parameters are fixed in a certain range, e.g., within each simulation subset, variations in the forcing wavenumber only weakly affect the normalization of the relation, rather than its functional form. However, it is worth mentioning that the simulations in subsets B, C, and D were all conducted on the same grid resolution. Thus, the associated error bars are expected to have comparable statistical properties, reflecting similar numerical errors and sampling uncertainties. Consequently, the absence of a statistically significant correlation should not be interpreted as definitive evidence for the absence of any $k_f$ dependence. Rather, resolving such a potentially weak trend would likely require either a broader survey of the parameter space in $k_f$, or simulations with reduced uncertainties, for example, through higher resolution or longer averaging times.

## 4. Conclusions

In the present work, we investigated the nondimensional coefficient $C_H$ in the magnetic energy spectrum of magnetically forced helical MHD. We numerically conducted 25 simulations varying multiple characteristic values of $\eta k_1/c_s$, $k_f/k_1$, $N^3$, $\text{Pr}_M$, and Lu. For each run, we scaled $\eta$ and observed a clear inverse cascade process in the magnetic energy spectrum. We then fitted $C_H$ using a systematic logarithmic sampling strategy and computed the compensated spectrum by $k\epsilon_H^{-2/3}$ to obtain an error estimation.

We extended our findings by investigating the Lu dependence of $C_H$ to the regime of high and low $\text{Pr}_M$ and multiple resolutions. Based on dimensional analysis, we tested two potential dependencies. Our results show that $C_H$ obeys an approximately linear dependence on Lu; the single-parameter fit for the coefficient is 0.24 with a coefficient of determination $R^2 = 0.82$. We also find that $C_H$ potentially obeys a power dependence on Lu with a power 2/3, and the single-parameter fit for the coefficient is 0.65 with a coefficient of determination $R^2 = 0.84$. This dependence is not affected by $\text{Pr}_M$ and $\eta$ in the current range investigated. Furthermore, we investigated the $k_f$ dependence of $C_H$ and found no clear statistical correlation between those two values.

For many astrophysical systems, the microscopic energy dissipation mechanism is not of Spitzer type, as assumed here, and the significance of Lu is unclear. It is not obvious how this would affect our results. It is probably true that a suitable value of Lu can be defined based on the growth rate of microphysical plasma instabilities. In any case, it is clear that conclusions based on $C_{\rm H}$ have a linear dependence on the Lundquist number.

Though it turns out that for large magnetic Prandtl numbers, most energy is dissipated viscously rather than resistively (A. Brandenburg 2014), a significant amount of energy could be dissipated resistively, especially when the magnetic energy strongly dominates over kinetic, for example in local accretion disk simulations (A. Brandenburg et al. 1995).

Our present work motivates possible avenues for future research. First, it highlights the significance of examining energy dissipation in astrophysical fluid dynamics, which is often ignored since most astrophysical fluid codes rely entirely on numerical prescriptions needed to dissipate energy when and where needed. In some extreme cases, for example, at very small values of $\mathrm{Pr_M}$, most of the energy is dissipated through resistivity rather than viscous dissipation, which fundamentally alters the energy cascade dynamics. While kinetic energy dissipation still occurs at small scales through viscous processes, the dominant energy dissipation pathway shifts to magnetic diffusion, making the inverse cascade in the magnetic energy spectrum a crucial mechanism that affects the overall energy dissipation in the system.

A critical verification requirement for describing the asymptotic regime is to confirm the independence of $C_{\rm H}$ from Lu across an extended range of parameter combinations. Given the inherent limitations imposed by finite numerical resolution, rectangular computational domains may still provide a viable approach to accessing a broader spectrum of spatial scales (A. Brandenburg et al. 2024). Additional strategies include implementing time-dependent profiles for $\eta$ and $\nu$ to achieve greater scaling with $\mathrm{Pr_M}$ and separation between forcing wavenumber $k_f$. However, such modifications introduce potential numerical artifacts that require rigorous validation. Care must be taken to distinguish physical phenomena from computational artifacts, particularly when employing hyperviscosity and hyperresistivity techniques, which are commonly utilized in MHD simulations, but may introduce poorly understood numerical effects that could compromise the physical interpretation of results.

Astrophysical systems, of course, have much larger values of Lu than what we have been able to simulate here. In the aforementioned paper by A. Brandenburg et al. (2024), for example, it was found that, by comparison with two-dimensional simulations, the dependence of another property, namely the ratio of the inverse cascade time over the Alfvén time, became eventually independent of Lu. In a similar way, it is plausible that our dependence of $C_{\rm H}$ on Lu could also eventually level off. However, showing this would require much higher resolution than what has been possible here.

Interestingly, our results suggest that, in contrast to $C_{\rm H}$, $C_{2/3}$ is nearly independent of Lu—even for small or moderately large values of Lu. Whether or not this lends support to the phenomenology of A. A. Ruzmaikin & A. M. Shukurov (1982) that the spectrum obeys the scaling $E_{\rm M}(k) \propto v_{\rm A}^2 k^{-1}$ is still unclear, because it makes no explicit reference to the presence of magnetic helicity. Indirectly, however, magnetic helicity is relevant for the very existence of the pronounced peak in the magnetic energy spectrum that travels toward progressively smaller $k$. The underlying field of scale $1/k$ could act as a large-scale magnetic field motivating the above phenomenology involving $v_{\rm A}$. Future work will hopefully offer more insight into the question of which of the possible phenomenologies is the relevant one to our system.

## Acknowledgments

We acknowledge the anonymous referee for making constructive suggestions that have improved the presentation of our paper. We also thank Giorgos Karydianakis for useful comments on our paper. This research was supported in part by the Swedish Research Council (Vetenskapsrådet) under grant No. 2019-04234, the National Science Foundation under grant Nos. NSF PHY-2309135, AST-2307698, and AST-2408411, and NASA Award 80NSSC22K0825. We acknowledge the allocation of computing resources provided by the Swedish National Allocations Committee at the Center for Parallel Computers at the Royal Institute of Technology in Stockholm.

## Data Availability

The source code used for the simulations of this study, the Pencil Code (Pencil Code Collaboration et al. 2021), is freely available on https://github.com/pencil-code. The simulation setups and corresponding input and reduced output data are freely available on doi:10.5281/zenodo.17940891.

## ORCID iDs

Jiyao Zhang https://orcid.org/0009-0006-0584-3818
Axel Brandenburg https://orcid.org/0000-0002-7304-021X